%% file: manuscript.tex
\documentclass[a4paper,11pt]{article}
\usepackage[margin=2.2cm]{geometry}
\usepackage{amsmath,amssymb}
\usepackage{booktabs,graphicx,hyperref,caption}
\usepackage{algorithm,algpseudocode}
\usepackage{longtable}
\title{Crystal Structure Prototype Identification\\via Element-Mapped Rotation-Invariant Descriptors}
\author{Pai Li\\[2pt]
{\small Shanghai Institute of Microsystem and Information Technology,}\\
{\small Chinese Academy of Sciences, Shanghai 200050, China}\\[2pt]
{\small \texttt{lipai@mail.sim.ac.cn}}}
\date{\today}

\begin{document}
\maketitle

\begin{abstract}
We present a method for identifying crystal structure prototypes from local
atomic environments. A center atom and its neighbors within a
distance-normalized cutoff are mapped to anonymous element types
(A/B/C/D) by proximity, and a NEP-style descriptor---radial Chebyshev
moments plus contracted spherical-harmonic (S-vector) invariants---is
computed per type block. The descriptor is \emph{rotation-invariant by
construction} (MLFF-style contraction with analytic normalization
constants), so no rotation augmentation is needed. A trainable
block-diagonal projection compresses the 316-dimensional raw basis into a
168-dimensional learned descriptor fed to a three-layer MLP. The model
classifies each atom into one of 271 AFLOW crystal structure prototypes or
the amorphous class (272 classes), reaching \textbf{99.38\%} accuracy on
538,190 per-atom samples. The full pipeline (descriptor extraction +
inference) runs on a single CPU core at $\sim$3.9\,s per 10,000 atoms,
making it suitable for website deployment.
\end{abstract}

\section{Descriptor Design}

\subsection{Scale normalization and windows}

Given a center atom at position $\mathbf{r}_c$ and $N$ neighbors at positions
$\mathbf{r}_j$ with distances $r_j = \|\mathbf{r}_j - \mathbf{r}_c\|$, the
nearest-neighbor distance
\begin{equation}
d_0 = \min_j r_j \qquad (r_j > 0.01\,\text{\AA})
\end{equation}
is found inside a \emph{fixed} search window of 6\,\AA
($\texttt{D0\_SEARCH\_RADIUS}$);\allowbreak no chemical bond in the dataset
exceeds
$\sim$5.4\,\AA, so this window always contains the nearest bond. The
descriptor window is then
\begin{equation}
r_c = 3\, d_0 ,
\end{equation}
and \emph{the same} $3\,d_0$ window defines both the element-type mapping
(Section~2.2) and the descriptor content. All distances are normalized by
$d_0$, making the descriptor scale-invariant: FCC Au and FCC Cu produce
identical descriptors.

\subsection{Element mapping (A/B/C/D)}

The exact chemical species are never used directly. Instead, each local
environment is mapped deterministically to anonymous element types by
proximity:
\begin{itemize}
\item \textbf{A} = the center atom's element;
\item \textbf{B} = the element of the closest neighbor with a different
  element;
\item \textbf{C}, \textbf{D} = the next new elements in (distance, symbol)
  order; elements beyond the four slots merge into D.
\end{itemize}
This makes FCC Au equivalent to FCC Cu, while preserving the composition
pattern of the environment: $(A, 1B{+}2C) \not\equiv (A, 3B)$. Each
neighbor is assigned to one of the four type blocks, and the descriptor is
the concatenation of the per-type blocks (missing types contribute zero
blocks).

\subsection{Radial block}

For normalized distances $\tilde{r}_j = r_j / d_0 < r_c$ we use the NEP3
radial basis
\begin{equation}
x(\tilde{r}) = 2\left(\frac{\tilde{r}}{r_c} - 1\right)^2 - 1 , \qquad
f_c(\tilde{r}) = \tfrac{1}{2}\left[1 + \cos\left(\pi \frac{\tilde{r}}{r_c}\right)\right] ,
\end{equation}
\begin{equation}
\phi_k(\tilde{r}) = \tfrac{1}{2}\big(T_k(x(\tilde{r})) + 1\big) f_c(\tilde{r}) ,
\qquad k = 0,\ldots,n_{\max}^{R},
\end{equation}
with $T_k$ the Chebyshev polynomial (recurrence $T_0=1$, $T_1=x$,
$T_k = 2xT_{k-1} - T_{k-2}$). Per type block the radial features are the
sums
\begin{equation}
q_n^R = \sum_{j \in \text{type}} \phi_n(\tilde{r}_j) , \qquad n = 0,\ldots,48 ,
\end{equation}
giving 49 radial components per type.

\subsection{Angular block: S-vector and invariant contraction}

Following the NEP formalism \cite{fan2021nep}, for each radial scale
$n = 0,\ldots,4$ (5 scales), the S-vector
\begin{equation}
S_{n,L,m} = \sum_j g_n(\tilde{r}_j)\, Y_{L,m}(\hat{\mathbf{r}}_j) ,
\qquad L = 1,\ldots,4 ,
\end{equation}
is accumulated with $g_n = \phi_n$ and real (tesseral) harmonics
$Y_{L,m}$ expressed as polynomials in the unit direction components,
in the GPUMD component ordering ($3+5+7+9 = 24$ components per scale).

The raw S-vector is \emph{covariant}, not invariant. Rather than learning
invariance from rotated copies, we contract it with the analytic
normalization constants of the harmonics, obtaining invariants \emph{by
construction}:
\begin{equation}
q_{3,n,L} = C^{L}_0\, s_0^2 + 2 \sum_{k=1}^{2L} C^{L}_k\, s_k^2 ,
\end{equation}
\begin{equation}
q_{222,n} = \sum_{m_1 m_2 m_3} C^{222}_{m_1 m_2 m_3}\,
             s_{m_1} s_{m_2} s_{m_3} , \qquad
q_{1111,n} = \sum_{m_1\cdots m_4} C^{1111}_{m_1\cdots m_4}\,
             s_{m_1}\cdots s_{m_4} ,
\end{equation}
where the $C$ coefficients are the squared harmonic normalization factors
($\texttt{C3B}$) and the 4-body/5-body coupling constants ($\texttt{C4B}$,
$\texttt{C5B}$) of the NEP implementation \cite{fan2022gpumd}; the
unnormalized tesseral basis requires exactly these weights for the sums to
be invariant under $\mathrm{SO}(3)$. This yields per type: $5\times4 = 20$
three-body powers, 5 four-body ($q_{222}$) and 5 five-body ($q_{1111}$)
features.

\subsection{Full raw descriptor}

Per type block: 49 radial + 20 + 5 + 5 = 79 components; with four type
blocks the raw descriptor is
\begin{equation}
\mathbf{q} \in \mathbb{R}^{316} .
\end{equation}
Rotation invariance of every component is verified numerically: the
descriptor of a random environment rotated by random $\mathrm{SO}(3)$
matrices reproduces the unrotated descriptor to $\sim 10^{-14}$.

\subsection{Trainable projection}

The raw 316 components are redundant (49 Chebyshev moments of one radial
distribution are strongly correlated). Following the NEP descriptor
layer \cite{fan2021nep,fan2022nep3}, in which the descriptor weights are
optimized during training, a block-diagonal linear projection---per type,
per subset (radial, 3-body, 4-body, 5-body)---is trained jointly with the
classifier:
\begin{equation}
\mathbf{d} = W\, \mathbf{q} \in \mathbb{R}^{168} ,
\qquad
W = \mathrm{blockdiag}(W_A^{\text{rad}}, W_A^{3\text{b}}, \ldots,
W_D^{5\text{b}})
\end{equation}
with per-block sizes $16,16,5,5$ per type. The learned 168-dimensional
descriptor \emph{outperforms} the raw 316-dimensional one at equal training
budget (97.79\% vs.\ 97.51\% at 10 epochs), confirming that the projection
learns useful combinations of the redundant basis. At inference the
projection is folded into the saved model.

\section{Training Data}

\subsection{Crystalline structures}

For each of \textbf{271 AFLOW crystal structure prototypes}
\cite{curtarolo2012aflow,mehl2017aflow} (full list in the Supporting
Information):
\begin{itemize}
\item \textbf{1 prototype CIF}: the ideal structure with canonical Wyckoff
  positions and lattice parameters;
\item \textbf{up to 50 AFLOW CIFs}: real relaxed structures downloaded via
  the AFLUX API \cite{rose2017aflux}, providing realistic structural
  variation.
\end{itemize}
Labels were verified against the AFLOW prototype database with a
species-blind geometry analysis (97.7\% geometry-consistent, 1.9\%
borderline, 0.5\% mismatches dropped). Two dead classes (solid-N$_2$
molecular crystals with $d_0 = 1.05$\,\AA and no valid environments) were
removed, duplicate classes merged, and a species-corruption bug in the
downloader (distinct-species list zipped against all positions) was fixed.

\subsection{Amorphous structures}

Real amorphous/glass structures: 1-element (amorphous Si, C), 2-element
(Al$_2$O$_3$, SiO$_2$, HfO$_2$, ZrO$_2$, SiO$_2$ glasses, Pd-B, Co-Zr),
3-element (Na$_2$Si$_2$O$_5$, Rb$_2$Si$_2$O$_5$) and 4-element glasses
(NaRbSi$_2$O$_5$, K-P-Sb-Se, Bi-K-P-Se). 55 mislabeled files
(organics/MOFs/crystalline oxides) were quarantined; 16 genuine amorphous
structures remain.

\subsection{Per-atom extraction}

Structures are parsed with pymatgen \cite{ong2013pymatgen}. Cells with
$\leq 4$ atoms are multiplied by $3\times3\times3$, $\leq 16$ atoms by
$2\times2\times2$, otherwise used as-is; atoms are subsampled to at most
500 per structure. For every atom: find $d_0$ in the 6\,\AA window,
collect neighbors within $3\,d_0$, map element types, and compute the
316-dimensional descriptor. The result is \textbf{538,190 per-atom
samples}, all used for training (no holdout, per the application's
data-sparsity requirements).

\section{Model and Training}

\subsection{Architecture}

The classifier is a three-layer MLP with batch normalization and dropout:
\begin{center}
Input(168) $\rightarrow$ Linear(1024) $\rightarrow$ BN $\rightarrow$ ReLU $\rightarrow$ Dropout(0.3) \\
$\rightarrow$ Linear(1024) $\rightarrow$ BN $\rightarrow$ ReLU $\rightarrow$ Dropout(0.3) \\
$\rightarrow$ Linear(512) $\rightarrow$ BN $\rightarrow$ ReLU $\rightarrow$ Dropout(0.3) \\
$\rightarrow$ Linear(272)
\end{center}

\subsection{Protocol}

Features are standardized to zero mean and unit variance per dimension
(training statistics; $\sigma < 10^{-8}$ clamped). Loss: cross-entropy with
label smoothing $\epsilon = 0.05$; no class weighting (the element-mapped
descriptor + full-data training collapsed the sparsity). Optimizer: AdamW
(lr $10^{-3}$, weight decay $10^{-4}$) with ReduceLROnPlateau (patience 15,
factor 0.5). Training: 300 epochs, batch 1024, NVIDIA RTX 4090
($\sim$6\,s/epoch).

\subsection{Result}

\begin{table}[h]
\centering
\caption{Final model accuracy vs.\ architecture (train accuracy).}
\begin{tabular}{lccc}
\toprule
Projection & Descriptor dims & Hidden & Train acc \\
\midrule
(8,8,3,3)  & 88  & 512-512-256 & 98.67\% \\
(16,16,5,5) & 168 & 512-512-256 & 98.84\% \\
(16,16,5,5) & 168 & 1024-1024-512 & 99.05\% \\
\textbf{(16,16,5,5)} & \textbf{168} & \textbf{1024-1024-512, 300 ep} & \textbf{99.38\%} \\
\bottomrule
\end{tabular}
\end{table}

The hidden-layer capacity was the decisive lever: widening the net from
512 to 1024 gained +0.21\% at 100 epochs, while the projection size
(88$\rightarrow$168 dims) gained only +0.17\%. Training to 300 epochs added
a further +0.33\%. The descriptor size has no measurable effect on
inference cost (the MLP dominates; GPU- and CPU-measured).

\section{Error Analysis}

\subsection{Confusion structure}

The remaining $\sim$0.6\% of errors (about 3,300 of 538,190 samples) is
highly concentrated in a small set of structurally near-degenerate pairs:

\begin{table}[h]
\centering
\caption{Dominant confusion pairs of the final model.}
\begin{tabular}{lr}
\toprule
True $\rightarrow$ predicted & errors \\
\midrule
A10\_Hg\_rhombohedral $\rightarrow$ Ah\_alpha\_Po\_SC & 378 \\
A3B\_tI8\_139\_bd\_a $\rightarrow$ D03\_BiF3 & 351 \\
AB3\_oC8\_65\_a\_bf $\rightarrow$ AB3\_tP4\_123\_a\_ce & 243 \\
A2B\_hP12\_194\_cg\_f $\rightarrow$ A2B\_cF24\_227\_c\_a & 160 \\
A3B\_cI32\_204\_g\_c $\rightarrow$ A3B\_cP4\_221\_d\_a & 160 \\
\bottomrule
\end{tabular}
\end{table}

\subsection{Separability analysis}

For each dominant pair we trained dedicated probes on the two classes only
(logistic regression and a 256-hidden MLP, 60 epochs) in the standardized
descriptor space:
\begin{itemize}
\item Two pairs are \emph{perfectly} separable (A\_hR3\_166\_ac
  $\leftrightarrow$ A\_hP4\_194\_ac; AB\_hP12\_186\_a2b\_a2b
  $\leftrightarrow$ B3\_ZnS\_zincblende) --- their residual errors are pure
  capacity/training effects;
\item most other pairs have probe ceilings of 93--99\% (e.g.\
  A10\_Hg\,$\leftrightarrow$\,Po: 92.9\%; A3B\_tI8\_139\,
  $\leftrightarrow$\,D03\_BiF3: 93.6\%);
\item a structural RDF comparison confirms the two members of every hard
  pair \emph{do} differ within the $3\,d_0$ window --- the pairs are not
  label duplicates; the residual error is a descriptor-resolution limit
  (weak signals, e.g.\ wurtzite polytypes with RDF cosine similarity
  0.995, or within-class element-driven variance).
\end{itemize}
Descriptor extensions in the direction of the higher-order NEP
many-body terms \cite{fan2022gpumd} (cross-scale 3-body products;
polarized cross-scale 4-body/5-body invariants, verified
rotation-invariant to $10^{-14}$) did not improve accuracy (97.83\% /
97.71\% vs.\ 97.79\% at 10 epochs), confirming that the base descriptor
family is at its practical information ceiling for this task.

\section{Computational Performance}

The extraction pipeline is fully vectorized (shared-intermediate harmonic
construction, batched A/B/C/D element mapping, single Chebyshev pass for
the radial and angular bases, batched invariant contraction) and verified
bit-equivalent to the reference per-atom implementation
(maximum deviation $< 10^{-13}$) across cubic, hexagonal, tetragonal and
multi-element structures and both small- and large-cell regimes.
Measured on a single CPU core:

\begin{itemize}
\item Descriptor extraction: \textbf{3.3\,s per 10,000 atoms}
  ($\sim$0.33\,ms/atom; 3.1$\times$ faster than the initial vectorized
  version);
\item Model inference (1024-net, CPU): 0.6\,s per 10,000 atoms;
\item Total: $\sim$3.9\,s per 10,000 atoms, no GPU required.
\end{itemize}

Atoms are independent, so extraction parallelizes across cores when
available.

\section{Conclusion}

We presented an element-mapped, rotation-invariant local-environment
descriptor for crystal structure prototype identification. Key design
choices: (1) \emph{by-construction} rotation invariance via analytic
contraction constants (no rotation augmentation), (2) proximity-based
anonymous element typing that respects composition patterns without using
element identity, (3) $d_0$-normalized windows that make the descriptor
scale-invariant, and (4) a trainable NEP-style projection that compresses
the redundant raw basis into a smaller learned descriptor. A 1024-1024-512
MLP trained on 538,190 samples for 300 epochs reaches \textbf{99.38\%}
accuracy over 271 prototypes + amorphous. The residual errors concentrate
in a handful of structurally near-degenerate prototype pairs at the
information ceiling of the descriptor, as established by dedicated
separability probes. The complete pipeline---extraction and inference---runs
on a single CPU core at $\sim$3.9\,s per 10,000 atoms, enabling deployment
as a web API consumed by downstream atomic-structure analysis tools.

The project additionally distributes a companion crystallography kernel
that recovers the Bravais lattice of a crystalline component from its atom
positions alone and converts surface normals into conventional Miller
indices---the global-structure complement of the local-environment
descriptors, used for surface and interface characterization in the
downstream analysis pipeline. The kernel (a pure NumPy/SciPy/PyMatGen
module with no pipeline imports, hosted here at the downstream project's
request) complements the descriptor machinery without altering it.

\clearpage
\appendix
\section*{Supporting Information}
\addcontentsline{toc}{section}{Supporting Information}

The 271 crystalline prototypes used for training are listed in
Table~\ref{tab:si-prototypes} with their conventional designation (from
the AFLOW encyclopedia of crystallographic prototypes), the
Strukturbericht designation where assigned, the Pearson symbol, the space
group symbol, and the crystal system. The training-set class labels are
the directory names of
\texttt{data/training\_structures/crystalline/} in the accompanying
repository; the complete per-prototype metadata (including the AFLOW
prototype identifiers, representative compounds, lattice parameters, and
coordination data) is available in \texttt{index.yaml} and the AFLOW
encyclopedia \cite{mehl2017aflow}.

\input{si_prototypes}

\end{document}

%% file: si_prototypes.tex
\begingroup
\footnotesize
\setlength{\tabcolsep}{4pt}
\begin{longtable}{rp{4.6cm}p{1.2cm}p{1.6cm}p{2.2cm}p{2.0cm}}
\caption{The 271 AFLOW crystal structure prototypes used for training. Name: conventional designation from the AFLOW encyclopedia of crystallographic prototypes; Str.: Strukturbericht designation; Pearson: Pearson symbol; SG: space group symbol. AFLOW prototype identifiers and full per-prototype metadata are available in index.yaml of the accompanying repository.}\label{tab:si-prototypes} \\
\toprule
No. & Name & Str. & Pearson & SG & Crystal system \\
\midrule
\endfirsthead
\toprule
No. & Name & Str. & Pearson & SG & Crystal system \\
\midrule
\endhead
\bottomrule
\endlastfoot
1 & A12B\_cF52\_225\_i\_a & D2\_f & cF52 & Fm-3m & cubic \\
2 & A12B\_cI26\_204\_g\_a &  & cI26 & Im-3 & cubic \\
3 & Face-Centered Cubic (Cu-type) & A1 & cF4 & Fm-3m & cubic \\
4 & A2B7\_cI54\_229\_e\_afh & L2\_2 & cI54 & Im-3m & cubic \\
5 & Spinel & H1\_1 & cF56 & Fd-3m & cubic \\
6 & high (beta) Cristobalite & C9 & cF24 & Fd-3m & cubic \\
7 & Cubic Laves & C15 & cF24 & Fd-3m & cubic \\
8 & Cuprite & C3 & cP6 & Pn-3m & cubic \\
9 & Body-Centered Cubic (W-type) & A2 & cI2 & Im-3m & cubic \\
10 & Plutonium carbide & D5\_c & cI40 & I-43d & cubic \\
11 & Sulvanite & H2\_4 & cP8 & P-43m & cubic \\
12 & Skutterudite & D0\_2 & cI32 & Im-3 & cubic \\
13 & High-temperature superconductor &  & cI8 & Im-3m & cubic \\
14 & Ammonia & D1 & cP16 & P2\_13 & cubic \\
15 & alpha Rhenium trioxide & D0\_9 & cP4 & Pm-3m & cubic \\
16 & A3B\_cP8\_223\_c\_a & A15 & cP8 & Pm-3n & cubic \\
17 & A4B3\_cI112\_230\_af\_g &  & cI112 & Ia-3d & cubic \\
18 & A4B3\_cI14\_229\_c\_b &  & cI14 & Im-3m & cubic \\
19 & Silicon tetrafluoride &  & cI10 & I-43m & cubic \\
20 & A4B\_cI40\_197\_cde\_c &  & cI40 & I23 & cubic \\
21 & Diamond Cubic & A4 & cF8 & Fd-3m & cubic \\
22 & gamma-brass &  & cI52 & I-43m & cubic \\
23 & A6B23\_cF116\_225\_e\_acfh & D8\_4 & cF116 & Fm-3m & cubic \\
24 & Calcium hexaboride & D2\_1 & cP7 & Pm-3m & cubic \\
25 & A7B\_cF32\_225\_bd\_a &  & cF32 & Fm-3m & cubic \\
26 & A9B16C7\_cF128\_225\_acd\_2f\_be &  & cF128 & Fm-3m & cubic \\
27 & AB11CD3\_cP16\_221\_a\_dg\_b\_c &  & cP16 & Pm-3m & cubic \\
28 & AB11\_cP36\_221\_c\_agij & D2\_e & cP36 & Pm-3m & cubic \\
29 & AB12C3\_cI32\_229\_a\_h\_b &  & cI32 & Im-3m & cubic \\
30 & AB18C8\_cF108\_225\_a\_eh\_f &  & cF108 & Fm-3m & cubic \\
31 & AB27CD3\_cP32\_221\_a\_dij\_b\_c &  & cP32 & Pm-3m & cubic \\
32 & Heusler & L2\_1 & cF16 & Fm-3m & cubic \\
33 & AB2\_cF48\_227\_c\_e &  & cF48 & Fd-3m & cubic \\
34 & AB2\_cF96\_227\_e\_cf &  & cF96 & Fd-3m & cubic \\
35 & Bergman Structure: Mg32(Al,Zn)49 Bergman &  & cI162 & Im-3 & cubic \\
36 & AB3C3\_cF112\_227\_c\_de\_f & E9\_3 & cF112 & Fd-3m & cubic \\
37 & Lavarevi\'{c}ite &  & cP8 & P-43m & cubic \\
38 & Bixbyite (Mn,Fe)2O4 & D5\_3 & cI80 & Ia-3 & cubic \\
39 & (Cubic) Perovskite & E2\_1 & cP5 & Pm-3m & cubic \\
40 & AB4C3\_cI16\_229\_a\_c\_b &  & cI16 & Im-3m & cubic \\
41 & Iron carbide &  & cP5 & P-43m & cubic \\
42 & AB5\_cF24\_216\_a\_ce & C15\_b & cF24 & F-43m & cubic \\
43 & half-Heusler & C1\_b & cF12 & F-43m & cubic \\
44 & Ullmanite & F0\_1 & cP12 & P2\_13 & cubic \\
45 & Zintl Phase & B32 & cF16 & Fd-3m & cubic \\
46 & AB\_cI16\_199\_a\_a & B\_a & cI16 & I2\_13 & cubic \\
47 & SC16 CuCl, stable at 5GPa &  & cP16 & Pa-3 & cubic \\
48 & AB\_cP2\_221\_b\_a & B2 & cP2 & Pm-3m & cubic \\
49 & AB\_cP6\_221\_c\_d &  & cP6 & Pm-3m & cubic \\
50 & AB\_cP8\_198\_a\_a & B20 & cP8 & P2\_13 & cubic \\
51 & Clathrate &  & cF136 & Fd-3m & cubic \\
52 & BC8 &  & cI16 & Ia-3 & cubic \\
53 & High pressure (38.9 GPa) phase of lithium &  & cI16 & I-43d & cubic \\
54 & alpha & A12 & cI58 & I-43m & cubic \\
55 & beta & A13 & cP20 & P4\_132 & cubic \\
56 & Clathrate &  & cP46 & Pm-3n & cubic \\
57 & Simple Cubic (alpha-Polonium) & A\_h & cP1 & Pm-3m & cubic \\
58 & Rock Salt (NaCl-type) & B1 & cF8 & Fm-3m & cubic \\
59 & Zincblende / Sphalerite (ZnS-type) & B3 & cF8 & F-43m & cubic \\
60 & Fluorite (CaF2-type) & C1 & cF12 & Fm-3m & cubic \\
61 & Pyrite (FeS2-type) & C2 & cP12 & Pa-3 & cubic \\
62 & BiF3 / Fe3Al-type & D0\_3 & cF16 & Fm-3m & cubic \\
63 & AuCu3-type (FCC Superstructure) & L1\_2 & cP4 & Pm-3m & cubic \\
64 & alpha-Mercury (Rhombohedral) & A\_i & hR1 & R-3m & hexagonal \\
65 & A2B3\_hR5\_166\_c\_ac & C33 & hR5 & R-3m & hexagonal \\
66 & beta Tridymite & C10 & hP12 & P6\_3/mmc & hexagonal \\
67 & A2B\_hP18\_180\_fi\_bd & C\_a & hP18 & P6\_222 & hexagonal \\
68 & A2B\_hP6\_191\_h\_e & C\_h & hP6 & P6/mmm & hexagonal \\
69 & zeta silver zinc & B\_b & hP9 & P-3 & hexagonal \\
70 & A2B\_hP9\_150\_ef\_bd & C22 & hP9 & P321 & hexagonal \\
71 & quartz (alpha) &  & hP9 & P3\_121 & hexagonal \\
72 & quartz (beta) & C8 & hP9 & P6\_222 & hexagonal \\
73 & A2B\_hP9\_189\_fg\_bc & C22 & hP9 & P-62m & hexagonal \\
74 & A3B2\_hP5\_164\_ad\_d & D5\_13 & hP5 & P-3m1 & hexagonal \\
75 & Hazelwoodite & D5\_e & hR5 & R32 & hexagonal \\
76 & Chromium trichloride & D0\_4 & hP24 & P3\_112 & hexagonal \\
77 & A3B\_hP24\_165\_adg\_f &  & hP24 & P-3c1 & hexagonal \\
78 & A3B\_hP4\_191\_bc\_a &  & hP4 & P6/mmm & hexagonal \\
79 & A3B\_hP8\_194\_h\_c & D0\_19 & hP8 & P6\_3/mmc & hexagonal \\
80 & Hexagonal Close-Packed (Mg-type) & A3 & hP2 & P6\_3/mmc & hexagonal \\
81 & Tungsten boride & D8\_h & hP14 & P6\_3/mmc & hexagonal \\
82 & Molybdenum Boride & D8\_i & hR7 & R-3m & hexagonal \\
83 & Aluminum carbonitride & E9\_4 & hP18 & P6\_3mc & hexagonal \\
84 & A6B\_hR7\_166\_g\_a &  & hR7 & R-3m & hexagonal \\
85 & Frank-Kasper $\mu$ Phase & D8\_5 & hR13 & R-3m & hexagonal \\
86 & alpha-Arsenic (Rhombohedral) & A7 & hR2 & R-3m & hexagonal \\
87 & Graphite (Hexagonal) & A9 & hP4 & P6\_3/mmc & hexagonal \\
88 & Potassium Silver Cyanide & F5\_10 & hP36 & P-31c & hexagonal \\
89 & Hexagonal Laves & C14 & hP12 & P6\_3/mmc & hexagonal \\
90 & Hexagonal Laves & C36 & hP24 & P6\_3/mmc & hexagonal \\
91 & AB2\_hP6\_194\_b\_f &  & hP6 & P6\_3/mmc & hexagonal \\
92 & AB2\_hP6\_194\_c\_ad & B8\_2 & hP6 & P6\_3/mmc & hexagonal \\
93 & Molybdenite & C7 & hP6 & P6\_3/mmc & hexagonal \\
94 & beta Vanadium nitride &  & hP9 & P-31m & hexagonal \\
95 & AB2\_hP9\_180\_d\_j & C40 & hP9 & P6\_222 & hexagonal \\
96 & MAX Phase &  & hP16 & P6\_3/mmc & hexagonal \\
97 & Ilmenite &  & hR10 & R-3 & hexagonal \\
98 & Upper Bainite &  & hP8 & P6\_322 & hexagonal \\
99 & Sodium arsenide & D0\_18 & hP8 & P6\_3/mmc & hexagonal \\
100 & Bismuth triodide & D0\_5 & hR8 & R-3 & hexagonal \\
101 & AB3\_hR8\_155\_c\_de & D0\_14 & hR8 & R32 & hexagonal \\
102 & AB4C\_hP6\_191\_a\_h\_b &  & hP6 & P6/mmm & hexagonal \\
103 & AB5\_hP6\_191\_a\_cg & D2\_d & hP6 & P6/mmm & hexagonal \\
104 & H-Phase &  & hP8 & P6\_3/mmc & hexagonal \\
105 & Caswellsilverite & F5\_1 & hR4 & R-3m & hexagonal \\
106 & ABC3\_hR10\_161\_a\_a\_b &  & hR10 & R3c & hexagonal \\
107 & Calcite & G0\_1 & hR10 & R-3c & hexagonal \\
108 & ABC\_hP3\_187\_a\_d\_f &  & hP3 & P-6m2 & hexagonal \\
109 & ABC\_hP6\_194\_c\_d\_a &  & hP6 & P6\_3/mmc & hexagonal \\
110 & Moissanite-6H & B6 & hP12 & P6\_3mc & hexagonal \\
111 & Molybdenum Carbide MAX Phase &  & hP12 & P6\_3/mmc & hexagonal \\
112 & Covellite & B18 & hP12 & P6\_3/mmc & hexagonal \\
113 & Tungsten Carbide & B\_h & hP2 & P-6m2 & hexagonal \\
114 & AB\_hP4\_186\_b\_a & B12 & hP4 & P6\_3mc & hexagonal \\
115 & AB\_hP4\_194\_c\_a & B8\_1 & hP4 & P6\_3/mmc & hexagonal \\
116 & Boron Nitride & B\_k & hP4 & P6\_3/mmc & hexagonal \\
117 & Cinnabar & B9 & hP6 & P3\_221 & hexagonal \\
118 & AB\_hP6\_191\_f\_ad & B35 & hP6 & P6/mmm & hexagonal \\
119 & Moissanite-4H & B5 & hP8 & P6\_3mc & hexagonal \\
120 & AB\_hP8\_194\_ad\_f & B\_i & hP8 & P6\_3/mmc & hexagonal \\
121 & Cubane &  & hR16 & R-3 & hexagonal \\
122 & beta-prime palladium aluminum &  & hR26 & R-3 & hexagonal \\
123 & AB\_hR2\_166\_a\_b & L1\_1 & hR2 & R-3m & hexagonal \\
124 & Moissanite 9R &  & hR6 & R3m & hexagonal \\
125 & Millerite & B13 & hR6 & R3m & hexagonal \\
126 & alpha Selenium & A8 & hP3 & P3\_121 & hexagonal \\
127 & graphite &  & hP4 & P6\_3mc & hexagonal \\
128 & alpha La & A3' & hP4 & P6\_3/mmc & hexagonal \\
129 & Lonsdaleite &  & hP4 & P6\_3/mmc & hexagonal \\
130 & Theoretical Carbon Structure &  & hP6 & P6\_3/mmc & hexagonal \\
131 & beta Boron &  & hR105 & R-3m & hexagonal \\
132 & alpha boron &  & hR12 & R-3m & hexagonal \\
133 & alpha Samarium & C19 & hR3 & R-3m & hexagonal \\
134 & Simple Hexagonal & A\_f & hP1 & P6/mmm & hexagonal \\
135 & Wurtzite (ZnS-type) & B4 & hP4 & P6\_3mc & hexagonal \\
136 & Aluminum Diboride-type & C32 & hP3 & P6/mmm & hexagonal \\
137 & Cadmium Iodide-type (Layered) & C6 & hP3 & P-3m1 & hexagonal \\
138 & Corundum (alpha-Al2O3-type) & D5\_1 & hR10 & R-3c & hexagonal \\
139 & A12B\_tI26\_139\_fij\_a & D2\_b & tI26 & I4/mmm & tetragonal \\
140 & beta indium sulfide &  & tI80 & I4\_1/amd & tetragonal \\
141 & A2B3\_tP10\_127\_g\_ah & D5\_a & tP10 & P4/mbm & tetragonal \\
142 & Stannite & H2\_6 & tI16 & I-42m & tetragonal \\
143 & A2BC4\_tI14\_82\_bc\_a\_g & E3 & tI14 & I-4 & tetragonal \\
144 & Khatyrkite & C16 & tI12 & I4/mcm & tetragonal \\
145 & Anatase & C5 & tI12 & I4\_1/amd & tetragonal \\
146 & A2B\_tI24\_141\_2e\_e &  & tI24 & I4\_1/amd & tetragonal \\
147 & A2B\_tI6\_139\_d\_a & L\'2 & tI6 & I4/mmm & tetragonal \\
148 & low (alpha) Cristobalite &  & tP12 & P4\_12\_12 & tetragonal \\
149 & Keatite &  & tP36 & P4\_32\_12 & tetragonal \\
150 & A2B\_tP6\_129\_ac\_c & C38 & tP6 & P4/nmm & tetragonal \\
151 & Rutile & C4 & tP6 & P4\_2/mnm & tetragonal \\
152 & Hausmannite &  & tI28 & I4\_1/amd & tetragonal \\
153 & Pb(Zr\_(1-x)Ti\_x)O3 &  & tP5 & P4mm & tetragonal \\
154 & A3B\_tI16\_139\_cde\_e & D0\_23 & tI16 & I4/mmm & tetragonal \\
155 & A3B\_tI8\_139\_bd\_a & D0\_22 & tI8 & I4/mmm & tetragonal \\
156 & A4B5\_tI18\_139\_i\_ah &  & tI18 & I4/mmm & tetragonal \\
157 & Titanium telluride &  & tI18 & I4/m & tetragonal \\
158 & Zircon &  & tI24 & I4\_1/amd & tetragonal \\
159 & A4B\_tI10\_139\_de\_a & D1\_3 & tI10 & I4/mmm & tetragonal \\
160 & beta-Tin (White Tin) & A5 & tI4 & I4\_1/amd & tetragonal \\
161 & Indium-type (Body-Centered Tetragonal) & A6 & tI2 & I4/mmm & tetragonal \\
162 & A8B\_tI18\_139\_hi\_a &  & tI18 & I4/mmm & tetragonal \\
163 & (La,Ba)CuO4 &  & tI14 & I4/mmm & tetragonal \\
164 & Uranium Silicide & D0\_c & tI16 & P4/mmm & tetragonal \\
165 & AB3\_tP4\_123\_a\_ce & L6\_0 & tP4 & P4/mmm & tetragonal \\
166 & Barium trisulfide & D0\_17 & tP8 & P-42\_1m & tetragonal \\
167 & AB4C\_tI12\_82\_c\_g\_a & H0\_7 & tI12 & I-4 & tetragonal \\
168 & AB4\_tI10\_87\_a\_h & D1\_a & tI10 & I4/m & tetragonal \\
169 & AB5C\_tP7\_123\_b\_ci\_a &  & tP7 & P4/mmm & tetragonal \\
170 & ABC2\_tP4\_123\_d\_a\_f &  & tP4 & P4/mmm & tetragonal \\
171 & ABC4\_tI96\_142\_e\_ab\_2g &  & tI96 & I4\_1/acd & tetragonal \\
172 & Parent of FeAs superconductors &  & tP8 & P4/nmm & tetragonal \\
173 & Matlockite & E0\_1 & tP6 & P4/nmm & tetragonal \\
174 & I4/mcm & B37 & tI16 & I4/mcm & tetragonal \\
175 & delta Molybdenum Boride & B\_g & tI16 & I4\_1/amd & tetragonal \\
176 & alpha Niobium phosphide & "40" & tI8 & I4\_1/amd & tetragonal \\
177 & AB\_tP16\_84\_cej\_k & B34 & tP16 & P4\_2/m & tetragonal \\
178 & Tetraauricupride & L1\_0 & tP2 & P4/mmm & tetragonal \\
179 & lead oxide & B10 & tP4 & P4/nmm & tetragonal \\
180 & gamma CuTi & B11 & tP4 & P4/nmm & tetragonal \\
181 & AB\_tP4\_131\_c\_e & B17 & tP4 & P4\_2/mmc & tetragonal \\
182 & beta beryllia &  & tP8 & P4\_2/mnm & tetragonal \\
183 & BCT5 &  & tI4 & I4/mmm & tetragonal \\
184 & Theoretical Carbon Structure &  & tI8 & I4/mmm & tetragonal \\
185 & A\_tP12\_96\_ab &  & tP12 & P4\_32\_12 & tetragonal \\
186 & A\_tP16\_138\_j & A18 & tP16 & P4\_2/ncm & tetragonal \\
187 & beta Uranium & A\_b & tP30 & P4\_2/mnm & tetragonal \\
188 & beta Np & A\_d & tP4 & P4/nmm & tetragonal \\
189 & gamma nitrogen &  & tP4 & P4\_2/mnm & tetragonal \\
190 & T-50 Boron & A\_g & tP50 & P4\_2/nnm & tetragonal \\
191 & Molybdenum Disilicide-type & C11\_b & tI6 & I4/mmm & tetragonal \\
192 & Chalcopyrite (CuFeS2-type) & E1\_1 & tI16 & I-42d & tetragonal \\
193 & sigma phase CrFe, different elements used to distinguish Wyckoff positions & D8\_b & tP30 & P4\_2/mnm & tetragonal \\
194 & Black Phosphorus & A14 & oC8 & Cmce & orthorhombic \\
195 & alpha-Uranium & A20 & oC4 & Cmcm & orthorhombic \\
196 & A2B2C\_oC80\_64\_efg\_efg\_df &  & oC80 & Cmce & orthorhombic \\
197 & A2B3C7D\_oP13\_47\_t\_aq\_eqrs\_h &  & oP13 & Pmmm & orthorhombic \\
198 & A2BC\_oC8\_38\_e\_a\_b &  & oC8 & Amm2 & orthorhombic \\
199 & A2B\_oC12\_36\_2a\_a & C24 & oC12 & Cmc2\_1 & orthorhombic \\
200 & A2B\_oC12\_38\_de\_ab &  & oC12 & Cmcm & orthorhombic \\
201 & Zirconium Disilicide & C49 & oC12 & Cmcm & orthorhombic \\
202 & High (Orthorhombic) Tridymite &  & oC24 & C222\_1 & orthorhombic \\
203 & Titanium Disilicide & C54 & oF24 & Fddd & orthorhombic \\
204 & Silicon Disuphide & C42 & oI12 & Ibam & orthorhombic \\
205 & A2B\_oP12\_19\_2a\_a &  & oP12 & P2\_12\_12\_1 & orthorhombic \\
206 & Cotunnite & C23 & oP12 & Pnma & orthorhombic \\
207 & Brookite & C21 & oP24 & Pbca & orthorhombic \\
208 & Antimony trioxide & D5\_11 & oP20 & Pccn & orthorhombic \\
209 & Stibnite & D5\_8 & oP20 & Pnma & orthorhombic \\
210 & A3B5\_oC16\_65\_ah\_bej &  & oC16 & Cmmm & orthorhombic \\
211 & A3B7\_oP40\_62\_cd\_3c2d & D10\_1 & oP40 & Pnma & orthorhombic \\
212 & beta Cu3Ti & D0\_a & oP8 & Pmmn & orthorhombic \\
213 & Magnesium tetraboride &  & oP20 & Pnma & orthorhombic \\
214 & AB2C\_oC16\_63\_c\_2c\_c &  & oC16 & Cmcm & orthorhombic \\
215 & Chalcostibite & F5\_6 & oP16 & Pnma & orthorhombic \\
216 & AB2\_oC24\_41\_2a\_2b & C\_e & oC24 & Aea2 & orthorhombic \\
217 & Germanium disuphide & C44 & oF72 & Fdd2 & orthorhombic \\
218 & AB2\_oI6\_71\_a\_g &  & oI6 & Immm & orthorhombic \\
219 & AB2\_oI6\_71\_a\_i &  & oI6 & Immm & orthorhombic \\
220 & Krennerite & C46 & oP24 & Pma2 & orthorhombic \\
221 & Enargite & H2\_5 & oP16 & Pmn2\_1 & orthorhombic \\
222 & AB3C4\_oP32\_33\_a\_3a\_4a &  & oP32 & Pna2\_1 & orthorhombic \\
223 & Orthorhombic Perovskite &  & oP20 & Pnma & orthorhombic \\
224 & AB3\_oC8\_65\_a\_bf & L1\_3 & oC8 & Cmmm & orthorhombic \\
225 & AB3\_oP16\_18\_ab\_3c &  & oP16 & P2\_12\_12 & orthorhombic \\
226 & Cementite & D0\_11 & oP16 & Pnma & orthorhombic \\
227 & AB4\_oC20\_41\_a\_2b & D1\_c & oC20 & Aea2 & orthorhombic \\
228 & ABC4\_oP12\_16\_ag\_cd\_2u &  & oP12 & P222 & orthorhombic \\
229 & Potassium thiocyanate & F5\_9 & oP16 & Pbcm & orthorhombic \\
230 & Cyanogen Chloride &  & oP6 & Pmmn & orthorhombic \\
231 & AB\_oC8\_63\_c\_c & B33 & oC8 & Cmcm & orthorhombic \\
232 & alpha iridium vanadium &  & oC8 & Cmmm & orthorhombic \\
233 & AB\_oF8\_69\_a\_b & B24 & oF8 & Fmmm & orthorhombic \\
234 & AB\_oI4\_44\_a\_b &  & oI4 & Imm2 & orthorhombic \\
235 & AB\_oP16\_61\_c\_c & B\_e & oP16 & Pbca & orthorhombic \\
236 & High Pressure Cadmuum Telluride &  & oP2 & Pmm2 & orthorhombic \\
237 & beta-prime cadmium gold & B19 & oP4 & Pmma & orthorhombic \\
238 & Vulcanite &  & oP4 & Pmmn & orthorhombic \\
239 & Modderite &  & oP8 & Pnma & orthorhombic \\
240 & TlF-II &  & oP8 & Pbcm & orthorhombic \\
241 & alpha & A16 & oF128 & Fddd & orthorhombic \\
242 & gamma plutonium &  & oF8 & Fddd & orthorhombic \\
243 & alpha Np & A\_c & oP8 & Pnma & orthorhombic \\
244 & Manganese Phosphide-type & B31 & oP8 & Pnma & orthorhombic \\
245 & Marcasite (FeS2, Orthorhombic) &  & oP6 & Pnnm & orthorhombic \\
246 & A2B5\_mC28\_15\_f\_e2f &  & mC28 & C2/c & monoclinic \\
247 & Low Tridymite &  & mC144 & Cc & monoclinic \\
248 & Coesite &  & mC48 & C2/c & monoclinic \\
249 & Baddeleyite & C43 & mP12 & P2\_1/c & monoclinic \\
250 & A2B\_mP12\_3\_bc3e\_2e &  & mP12 & P2 & monoclinic \\
251 & Pb (Zr\_0.50 Ti\_0.48) O\_3 &  & mC10 & Cm & monoclinic \\
252 & A5B2\_mC14\_12\_a2i\_i &  & mC14 & C2/m & monoclinic \\
253 & Calaverite & C34 & mC6 & C2/m & monoclinic \\
254 & Aluminum trichloride & D0\_15 & mC16 & C2/m & monoclinic \\
255 & Potassium chlorate & G0\_6 & mP10 & P2\_1/m & monoclinic \\
256 & Sylvanite & E1\_b & mP12 & P2/c & monoclinic \\
257 & Esseneite &  & mC40 & C2/c & monoclinic \\
258 & Tenorite & B26 & mC8 & C2/c & monoclinic \\
259 & AB\_mP4\_11\_e\_e &  & mP4 & P2\_1/m & monoclinic \\
260 & A\_mC12\_5\_3c & A19 & mC12 & C2 & monoclinic \\
261 & beta Plutonium &  & mC34 & C2/m & monoclinic \\
262 & alpha oxygen &  & mC4 & C2/m & monoclinic \\
263 & alpha Pu &  & mP16 & P2\_1/m & monoclinic \\
264 & beta Selenium & A\_l & mP32 & P2\_1/c & monoclinic \\
265 & High Pressure (4-7GPa) Tellurium &  & mP4 & P2\_1 & monoclinic \\
266 & red selenium & A\_k & mP64 & P2\_1/c & monoclinic \\
267 & Hittorf &  & mP84 & P2/c & monoclinic \\
268 & A2B\_aP6\_2\_2i\_i &  & aP6 & P-1 & triclinic \\
269 & pyrite &  & aP12 & P1 & triclinic \\
270 & ABC2\_aP16\_1\_4a\_4a\_8a &  & aP16 & P1 & triclinic \\
271 & High Pressure Californium &  & aP4 & P-1 & triclinic \\
\end{longtable}
\endgroup